\documentstyle[twocolumn,prl,aps]{revtex}
\begin{document}

\wideabs{

\title{Exclusion Statistics of Quasiparticles in 
Condensed States of Composite Fermion 
Excitations}

\author{Piotr Sitko} 

\address{Institute of Physics, 
Wroc\l{}aw University of Technology,\\ Wybrze\.ze Wyspia\'{n}skiego
 27, 50-370 Wroc\l{}aw, Poland.}

\maketitle

\begin{abstract}
The   exclusion statistics 
of quasiparticles 
is found at any level
of the hierarchy of condensed states of composite fermion excitations
(for which  experimental indications have recently been found).
The hierarchy of condensed states 
 of excitations in boson  
Jain states is  introduced
and the statistics of quasiparticles  is found.
The  quantum Hall states  
of charged $\alpha$-anyons ($\alpha$ -- the exclusion statistics
parameter) can be described as incompressible states
of $(\alpha+2p)$-anyons ($2p$ -- an even number).

\end{abstract}

}



Recent fractional quantum Hall experiments by Pan {\it et al.}
\cite{Pan} (described  in \cite{Smet})
indicate the presence of the new odd-denominator hierachical states
(other than Laughlin and Jain states), also indications of
even-denominator hierarchical states \cite{Pan,filling} are found.
If  the results are confirmed they will bring a  breaktrough
in our knowledge of
the fractional quantum Hall effect.
The hierarchy of odd-denominator states in the fractional quantum Hall
effect was introduced by Haldane at early stages of the quantum Hall
theory 
\cite{Haldane1} (short time 
after  Laughlin presented his original approach \cite{Laughlin}).
Within years, however, the experiments  supported
only  hierarchical
states given by the composite fermion approach.
The composite fermion approach predicts  the so-called Jain states 
at  filling fractions of
the form $\nu=\frac{n_{0}}{2p_{0}n_{0}+\beta_{0}}$ 
(where $2p_{0}$ -- the Chern-Simons composite fermion parameter being an even
number,
$n_{0}$ -- the number of effective shells filled,
$\beta_{0}$ -- the sign of the effective 
field with respect to the
external magnetic field).
The hierachical theory of condensed states of 
composite fermion excitations was also proposed
(in a close analogy with the Haldane hierarchy),
but  hierachical states were not numerically verified \cite{QHS}
(e. g.  the $4/11$
 state -- for which the strongest evidence is found \cite{Pan}).
The experiment \cite{Pan} opens the possibility for a realization of
condensed states of composite fermion excitations \cite{cfhierarchy}.

The anyon statistics parameters of quantum Hall quasiparticles
were determined by Halperin within the Haldane hierarchy picture
\cite{Halperin}.
Here, we present the analogous approach, however, the quasiparticle statistics
is determined within the generalized exclusion statistics  approach for
condensed states  of  excitations in Jain states
 on a sphere.
In Laughlin and Jain fermion states  quasiparticles obey
fractional statistics\cite{Halperin,Arovas} with exclusion statistics 
parameters  defined  earlier
\cite{He,Johnson,Isakov,Sitko2002,Su}.
Let us add that recently  it was 
proposed  that quantum Hall quasiparticles obeying
fractional statistics may serve as a good tool for quantum computation
\cite {Averin} (called topological quantum computation \cite{Zanardi}).

In this paper we will determine the statistics of quasiparticles
in hierarchical states using the composite fermion approach
(and folowing the construction of the 
hierarchy of condensed states of composite
fermion excitations \cite{cfhierarchy}).
We make use of the generalized
Haldane  exclusion statistics \cite{Haldane2,Wu} (with the statistics
parameter $\alpha$).
The anyon statistics parameter  of   quantum Hall quasiparticles  
 is rederived as 
$\frac{\theta}{\pi}=\alpha(mod\; 2)$  
-- see \cite{Wu,Ouvry} ($\theta$ is the phase change which appears in the
wave function when two anyons are interchanged).

The generalized exclusion statistics for 
identical particles can be expressed in terms of the
number of many-particle states as \cite{Wu}:
\begin{equation}
\label{def}
k+(1-\alpha)(N-1) \choose N
\end{equation}
where $k$ is the number of available
single-particle states  and
$\alpha$ is the statistics parameter, $N$ -- the number of particles.
So that, we are going to define the number of many-particle states for
quasiparticles in hierarchical  states using the composite fermion
approach for a spherical system \cite{Chen} (for
excitations in Laughlin  and Jain states
 it was done in \cite{He,Johnson,Isakov,Sitko2002,Su}).
The quasiparticles in even-denominator quantum Hall 
states are predicted to obey
nonabelian fractional statistics \cite{QHS,Read} and are not considered
here.

The composite fermion approach was extensively  and successfully  tested
in many numerical studies of quantum Hall systems  -- i. e.
of  systems of electrons in a magnetic field 
such that  the lowest Landau level is  partially filled (see for example
\cite{QHS}).
The numerical results are available only for 
a small number of electrons
(usually up to 12-14 electrons) and there are  only 
data for a few quasiparticles \cite{cfhierarchy}.
In fact,  the $4/11$ state is not predicted in numerical
studies for Coulomb interaction \cite{QHS} (it is important
to add that the experiment \cite{Pan} indicates
 that the observed $4/11$ state is  fully spin polarized).

The composite fermion hierarchy consists in the assumption that
quasiparticles  (e. g. in Laughlin states)
can be treated in the same way as electrons (partially
filling the lowest Landau level) are treated.
So that, we can use the composite fermion approach (well
established for
electrons partially filling the lowest Landau level \cite{QHS}) 
to quasiparticles partially filling a shell
(all  predictions will be the same as for electrons).

Let us introduce the hierarchy of condensed states of composite fermion
excitations (following \cite{cfhierarchy}).
The hierarchy fractions is defined by the set of equations:
\begin{equation}
\label{fraction}
\nu_{0}^{-1}=2p_{0}+\frac{\beta_{0}}{n_{0}+\nu_{1}}
\end{equation}
where $\nu_{1}$ is the fraction in which the next $(n_{0}+1)$-th 
effective shell is filled.
The procedure (\ref{fraction})  can be repeated on $\nu_{1}$,
and so on, to find the final odd-denominator filling fraction:
\begin{equation}
\label{hierarchy}
\nu_{0}^{-1}=2p_{0}+\frac{\beta_{0}}{n_{0}+
...\; +\frac{1}{2p_{k}+\frac{\beta_{k}}{n_{k}} }     }\; .
\end{equation}
The hierarchy  of even denominator fractions is found when $\beta_{k}=0$ at
any step of the hierarchy \cite{filling}.
The quantum Hall states at those fractions would be related to Pfaffian
states of quasiparticles (see for example \cite{Read,QHS}).
Still, in order to define the Pfaffian states one needs the statistics of
quasiparticles.


In the case of a spherical system  the composite fermion effective field
is given by \cite{QHS}:
\begin{equation}
2S^{*}=\beta_{0}(2S-2p_{0}(N_{e}-1))\; 
\end{equation}
where $2S$ is the number of flux quanta piercing the sphere, 
 $N_{e}$ --  the number of electrons, $2S^{*}$ -- the effective composite fermion field,
$\beta_{0}$ is the sign of the effective field with respect to $2S$.
One has also the relations:
$2S_{qe(0)}=2S^{*}+2n_{0}$, $2S^{*}n_{0}+n_{0}^{2}=N_{e}-N_{qe(0)}$ 
(if only
quasielectrons are present).

The number of many-particle states (at any step of
the hierarchy) is given by:
\begin{equation}
{2S_{qe} +1 \choose N_{qe}} \; .
\end{equation}
One can also perform the composite fermion transformation for this effective
shell
\begin{equation}
2S_{qe}^{*}=\beta_{qe}(2S_{qe}-2p_{qe}(N_{qe}-1))\; ,
\end{equation}
$\beta_{qe}$ is the sign of the effective quasiparticle shell
(with respect to $2S_{qe}$).

Let us first find the statistics parameters 
of quasiparticles in Jain states 
\cite{Isakov,Sitko2002}, and we will determine the statistics of
quasiparticles in the next step of the hierarchy.
Using the definition (1) one can write
\begin{equation}
\label{pocz}
2S_{qe(0)}= (1-\alpha_{qe(0)})N_{qe(0)}+A(2S,n_{0},2p_{0},\beta_{0})
\end{equation}
where $A$ is a function of $2S$, $n_{0}$, $2p_{0}$, $\beta_{0}$
\cite{Isakov,Sitko2002}:
\[
A=\beta_{0}2S +\beta_{0}2p_{0}+2n_{0}-
\beta_{0}2p_{0}\frac{\beta_{0}n_{0}(2S+2p_{0})+n_{0}^{2}}{1+2p_{0}
\beta_{0}n_{0}}\; ,
\]
and
$
\label{Jain}
1-\alpha_{qe(0)}=\frac{-\beta_{0}2p_{0}}{1+2p_{0}\beta_{0}n_{0}}\; 
$
(as was found in \cite{Isakov,Sitko2002}).
One gets next
\begin{equation}
2S_{qe(0)}^{*}=\beta_{1}(2S_{qe(0)}-2p_{1}(N_{qe(0)}-1))\; ,
\end{equation}
and
$
2S_{qe(1)}=2S_{qe(0)}^{*}+2n_{1}\; 
$.
The number of many particle states for $N_{qe(1)}$ is given by
${2S_{qe(1)}+1\choose N_{qe(1)}} \;$.
We need also the relation (if only quasielectrons are present):
$
N_{qe(0)}-N_{qe(1)}=2S_{qe(0)}^{*}n_{1}+n_{1}^{2}
$,
(or if only quasiholes are present)
$
N_{qe(0)}+N_{qh(1)}=2S_{qe(0)}^{*}n_{1}+n_{1}^{2}\; 
$.
Note that here 
we perform the composite fermion transformation only on
quasielectrons (not on quasiholes -- we use the $\beta=-1$ effective field
instead \cite{filling}).
The final result is:
\begin{equation}
\frac{1}{1-\alpha_{qe(1)}}=
\frac{1}{\beta_{1}(1-\alpha_{qe(0)} -2p_{1})}-n_{1}=
\frac{1}{\alpha_{qh(1)}-1}
\end{equation}
We can also write the general relation  (for quasiparticles at the
$k$-step of the hierachy):
\begin{equation}
\label{relation}
\frac{1}{1-\alpha_{qe(k)}}=
\frac{1}{\beta_{k}(1-\alpha_{qe(k-1)} -2p_{k})}-n_{k}\; ,
\end{equation}
and $\alpha_{qh(k)}-1=1-\alpha_{qe(k)}$.
One can notice that the statistics of composite fermion excitations in
Jain states 
can be
found by putting $\alpha_{qe(-1)}=1$ (simply because
electrons are fermions).
It can be rewritten as
$
\alpha_{qe(0)}=1+\frac{\beta_{0}2p_{0}}{1+\beta_{0}n_{0}2p_{0}}
$
and for quasiholes in Jain states
$
\alpha_{qh(0)}=1-\frac{\beta_{0}2p_{0}}{1+\beta_{0}n_{0}2p_{0}}
$
in agreement with previous results \cite{Isakov,Sitko2002}.

We can define the statistics parameter at any level of the
hierarchy by the set of  equations 
\begin{displaymath}
\frac{1}{1-\alpha_{qe(0)}}=
\frac{1}{-\beta_{0}2p_{0}}-n_{0}
\end{displaymath}
\begin{displaymath}
\frac{1}{1-\alpha_{qe(1)}}=
\frac{1}{\beta_{1}(1-\alpha_{qe(0)} -2p_{1})}-n_{1}
\end{displaymath}
\begin{displaymath}
\frac{1}{1-\alpha_{qe(2)}}=
\frac{1}{\beta_{2}(1-\alpha_{qe(1)} -2p_{2})}-n_{2}
\end{displaymath}
\begin{displaymath}
...
\end{displaymath}
\begin{equation}
\label{set}
\frac{1}{1-\alpha_{qe(k)}}=
\frac{1}{\beta_{k}(1-\alpha_{qe(k-1)} -2p_{k})}-n_{k}
\end{equation}
and $1-\alpha_{qe(k)}=\alpha_{qh(k)}-1$.

The Haldane hierachy corresponds to the choice
$n_{k}=1$ at every step of the
hierarchy (\ref{hierarchy})
\cite{cfhierarchy}, then one can find the result
(which is analogous to the Eq. (4) of Ref. \cite{Halperin}):
\begin{equation}
\label{Halp}
\frac{\beta_{k}}{\alpha_{qh(k)}}=(2p_{k}+1+\beta_{k})-
\alpha_{qh(k-1)} \;.
\end{equation}
Note again that we use the $\beta=-1$ effective field instead of
considering condensed states of quasiholes.
One can easily notice that the denominator in $\alpha_{qh(k-1)}$
is the numerator in $\alpha_{qh(k)}$ (for the Haldane hierarchy).
The Jain sequence within the Haldane hierarchy
is found for $p_{k}=0$ and $n_{k}=1$, $\beta_{k}=1$, at any step.

Let us consider the second order of the hierarchy.
For the $4/11$ state ($p_{0}=1$, $n_{0}=1$, $\beta_{0}=1$,
$p_{1}=1$, $n_{1}=1$, $\beta_{1}=1$) one finds
$\alpha_{qh(1)}(\nu=4/11)=
\frac{3}{11}$ and $\alpha_{qe(1)}(\nu=4/11)=2-\frac{3}{11}$.
For several other  fractions the statistics parameters
are given in Table I. One can observe that (in agreement with  results of
Halperin \cite{Halperin}) $\alpha_{qh(k)}=p/q$ where $p$ is the
denominator of the parent state and $q$ is the denominator of the daughter
state. For $\nu_{0}=5/13$ 
the parent state (in the spirit of the Haldane hierarchy)
is the $2/5$ state (to get it one can put $n_{1}=1$ in the second row
of Table I),
for $\nu_{0}=5/17$  the parent state is the $2/7$ state ($n_{1}=1$).
The $31/65$ state is the daughter state of the Jain $10/21$ state
\cite{Pan}.

Since we perform the composite fermion transformation only on
quasielectrons  the  $4/13$ state
is found with $p_{0}=2$ and  $\beta_{0}=-1$  ($n_{0}=1$)
for the Laughlin $1/3$
state. One can also put $p_{0}=1$, $\beta_{0}=1$, $n_{0}=1$, to get the
Laughlin $1/3$ state within the composite fermion approach.
The  excitations above the $1/3$ state  can also be described in two ways
(both descriptions can be to some extent
verified numerically -- see for example \cite{Yi}).
For example  the $2/7$ state can be seen as the $1/3$ state of
quasiholes (within the $p_{0}=1$ description) or,
with $p_{0}=2$, it is the Jain state
(first order of the hierachy, or $\nu_{1}=1$ -- quasielectrons
completely fill the shell $2S_{qe}$).
It is apparent that  numbers of many-particle states for 
quasiparticles in the two descriptions differ \cite{Yi}.
The lowest energy states within the $p_{0}=1$ description
can be seen as states of quasielectrons  within the $p_{0}=2$ description
\cite{Yi}.
Specifically at $\nu_{0}=2/7$ one gets the single many-particle state  of
($p_{0}=2$) quasielectrons  (the ground state)   and
$3(N_{qh(0)}-1)+1\choose N_{qh(0)}$ many-particle states of ($p_{0}=1$)
quasiholes.
Of course, when one would perform the composite fermion transformation
on those  quasiholes one would find the single-many particle state
(the ground state at $\nu_{0}=2/7$).
We argue then that different    descriptions of excited states are 
related through the
composite fermion transformation.
Quasiparticles within different descriptions are related by
 (we consider
 quasiholes ($p_{k}=p$)
and corresponding quasielectrons ($p_{k}=p+1$)):
\[
\alpha_{qe(k)}^{(p_{k}=p+1,\beta_{k}=-1,n_{k}=1)}-
\alpha_{qh(k)}^{(p_{k}=p,\beta_{k}=1, n_{k}=1)}=2\; 
\]
which exactly corresponds to  the composite fermion 
transformation on quasiparticles.
For example 
Laughlin quasiholes ($p_{0}=1$, $\alpha_{qh(0)}=1/3$)
 correspond to quasielectrons with $p_{0}=2$ and $\alpha_{qe(0)}=7/3$.
The anyon statistics parameter 
within different descriptions equals
$\theta/\pi=\alpha (mod \; 2)$.
In general the composite fermion transformation on
quasiparticles ($\alpha$-anyons)
would give "superanyons" of statistics $\alpha +2p$ ($2p$
is an even number).
This
corresponds to the description of
the system of electrons
in terms of composite fermions \cite{Wu,Ouvry,Sitko2002} 
(called "superelectrons"
by Wu \cite{Wu}).

There is a growing interest in studies of quantum Hall boson systems
(expected to be found in rotating Bose gases \cite{Wilkin}).
Below we define the hierarchical  states in boson quantum Hall systems
for which $\nu_{F}^{-1}$=$\nu_{B}^{-1}+1$ 
($\nu_{F}$ is the filling fraction for fermions,
$\nu_{B}$ -- the corresponding filling fraction for bosons \cite{Xie}).
We do not consider here the Pfaffian boson states 
\cite{Wilkin} (or Read and Reazyi
states \cite{Read,Wilkin}).
Within the above approach we only have to change $2p_{0}$ into
$2p_{0}-1$ and other relations in (\ref{set})
remain the same.
For the Jain boson states one finds:
\begin{equation}
\frac{1}{1-\alpha_{qe(0)}}=
\frac{1}{\beta_{0}(1-2p_{0})}-n_{0} =
\frac{1}{\alpha_{qh(0)}-1} \; .
\end{equation}
For example at $\nu_{B}=\frac{1}{2}$,
$\alpha_{qe(0)}(\nu_{B}=\frac{1}{2})=\frac{3}{2}$,
$\alpha_{qh(0)}(\nu_{B}=\frac{1}{2})=\frac{1}{2}$.
For the hierarchical state (the condensed state of excitations in the
Laughlin
boson state $\nu_{B}=1/2$) $\nu_{B}=4/7$ (corresponding to the
$\nu_{F}=4/11$ fermion state) one gets
$\alpha_{qh(1)}(\nu_{B}=4/7)=\frac{2}{7}$.  For several other hierachical
boson fractions the statistics of quasiparticles is given in Table II.
The parent state for the boson fillings
$4/7$, $6/11$, $4/9$ (in the spirit of the Haldane
hierarchy) is the Laughlin $1/2$ boson state.
The parent states for $5/8$ and $5/12$  are  $2/3$ and $2/5$,
respectively.
For $\nu_{B}=31/34$ the parent state is $\nu_{B}=10/11$.

We want to underline that the relation (\ref{relation})
satisfies the 
symmetry 
$
QE(2p_{0}, \beta_{0}, n_{0}, \; ...,2p_{k},\beta_{k}, n_{k})-
$
$
QH(2p_{0}, \beta_{0}, n_{0}, \; ...,2p_{k},\beta_{k}, n_{k}+1)
$
(all parameters in parentheses are the same except for $n_{k}$,
for bosons $2p_{0}\longrightarrow (2p_{0}-1)$):
\begin{equation}
\label{duality}
\alpha_{qe}(n_{k})=\alpha_{qh}^{-1}(n_{k}+1)\; .
\end{equation}
which is the result of the  particle-hole conjugation  (quasielectrons and quasiholes are in the
same effective shell) -- this corresponds to the duality
described by Nayak and Wilczek \cite{Nayak,Sitko2002}.
The Eq. (\ref{duality}) is identical with the Eq. (\ref{Halp}) for
$p_{k}=0$ and $\beta_{k}=1$.
Also the equation (\ref{duality}) is valid for the special case
when $n_{k}=0$. For $k=0$ one finds 
$\alpha_{qe}(n_{0}=0)=2p_{0}+1$ ($\beta_{0}=1$)
-- the exclusion statistics of composite fermions \cite{Wu,Ouvry}.
For $k>0$ one finds
"superanyons" -- for example
quasielectrons conjugated to quasiholes with the exclusion statistics
parameters given in Tables (the last column).
At filling fractions given in the first column 
of the Tables these "superanyons"
 fill a single effective shell (they form incompressible states).


In conclusion, we 
found the exclusion 
statistics parameter  $\alpha$ for quasiparticles
in odd-denominator 
hierarchical condensed 
states of composite fermion excitations on a sphere.
The statistics parameters for quasiparticles in
the hierarchy of condensed  states of excitations in Jain
boson states  are also found.
The statistics parameters $\alpha$
obey quasiparticle -- quasihole 
 symmetry (when they are in the same effective shell). 
The anyon statisics parameter for quasiparticles 
is obtained as $\frac{\theta}{\pi}=\alpha
(mod\; 2)$.  
The quantum Hall states of  charged anyons
of the exclusion statistics parameter $\alpha$ 
can be described as  incompressible states
of $(\alpha+2p$)-anyons.


This work was supported by KBN grant.




\begin{table}[t]
\begin{center}
\begin{tabular}{|c|c|c|c|c|c|c|c|c|}
$\nu^{F}_{0}$& $p_{0}$& $n_{0}$& $\beta_{0}$& $\alpha_{qe(0)}$&
$p_{1}$& $n_{1}$ & $\beta_{1}$ & $\alpha_{qh(1)}$\\
\hline
$4/11$& $1$ & $1$& $1$& $5/3$& $1$& $1$& $1$& $3/11$\\
\hline
$5/13$& $1$ & $1$& $1$& $5/3$& $1$& $2$& $-1$& $5/13$\\
\hline
$6/17$& $1$ & $1$& $1$& $5/3$& $2$& $1$& $1$& $3/17$\\
\hline
$4/13$& $2$ & $1$& $-1$& $7/3$& $1$& $1$& $1$& $3/13$\\
\hline
$5/17$& $2$ & $1$& $-1$& $7/3$& $1$& $2$& $-1$& $7/17$\\
\hline
$31/65$& $1$ & $10$& $1$& $23/21$& $1$& $1$& $1$& $21/65$\\
\end{tabular}
\end{center}
\caption{The exclusion statistics parameters for several filling
fractions for fermion quantum Hall systems.} 
\end{table}

\begin{table}[t]
\begin{center}
\begin{tabular}{|c|c|c|c|c|c|c|c|c|}
$\nu^{B}_{0}$& $2p_{0}-1$& $n_{0}$& $\beta_{0}$& $\alpha_{qe(0)}$&
$p_{1}$& $n_{1}$ & $\beta_{1}$ & $\alpha_{qh(1)}$\\
\hline
$4/7$& $1$ & $1$& $1$& $3/2$& $1$& $1$& $1$& $2/7$\\
\hline
$5/8$& $1$ & $1$& $1$& $3/2$& $1$& $2$& $-1$& $3/8$\\
\hline
$6/11$& $1$ & $1$& $1$& $3/2$& $2$& $1$& $1$& $2/11$\\
\hline
$4/9$& $3$ & $1$& $-1$& $5/2$& $1$& $1$& $1$& $2/9$\\
\hline
$5/12$& $3$ & $1$& $-1$& $5/2$& $1$& $2$& $-1$& $5/12$\\
\hline
$31/34$& $1$ & $10$& $1$& $12/11$& $1$& $1$& $1$& $11/34$\\
\end{tabular}
\end{center}
\caption{The exclusion statistics parameters for several filling
fractions for boson quantum Hall systems.} 
\end{table}

\end{document}